\documentclass[runningheads]{llncs}
\usepackage{graphicx}
\usepackage{algorithm}
\usepackage{algorithmic}
\usepackage{amsmath}
\usepackage{amssymb}
\usepackage{graphicx}
\usepackage{array}
\usepackage{booktabs}
\usepackage{balance}
 \usepackage{mathtools}
\usepackage{wrapfig}

\usepackage{caption}
\usepackage{subcaption}
\usepackage{multirow}
\usepackage{multicol}
\usepackage{float}
\usepackage{enumerate}
\usepackage[shortlabels]{enumitem}
 \usepackage{color,soul}
\usepackage{tabularx}

\begin{document}
\title{Estimating Topic Exposure for Under-Represented Users on Social Media}
%
%
\author{Mansooreh Karami\orcidID{0000-0002-8168-8075} \and \\
Ahmadreza Mosallanezhad\orcidID{0000-0003-1907-3536} \and \\
Paras Sheth\orcidID{0000-0002-6186-6946} \and \\
Huan Liu\orcidID{0000-0002-3264-7904}}
%
\institute{Arizona State University, Tempe AZ, USA
\email{\{mkarami,amosalla,psheth5,huanliu\}@asu.edu}}
\maketitle              
\begin{abstract}
Online Social Networks (OSNs) facilitate access to a variety of data allowing researchers to analyze users' behavior and develop user behavioral analysis models. These models rely heavily on the observed data which is usually biased due to the participation inequality. This inequality consists of three groups of online users: the lurkers - users that solely consume the content, the engagers - users that contribute little to the content creation, and the contributors - users that are responsible for creating the majority of the online content. Failing to consider the contribution of all the groups while interpreting population-level interests or sentiments may yield biased results. To reduce the bias induced by the contributors, in this work, we focus on highlighting the engagers' contributions in the observed data as they are more likely to contribute when compared to lurkers, and they comprise a bigger population as compared to the contributors. The first step in behavioral analysis of these users is to find the topics they are exposed to but did not engage with. To do so, we propose a novel framework that aids in identifying these users and estimates their topic exposure. The exposure estimation mechanism is modeled by 
incorporating behavioral patterns from similar contributors as well as users' demographic and profile information.

\end{abstract}
\section{Introduction}
In the modern era, Online Social Networks (OSNs) have been abundantly available and easily accessible worldwide. Thus, OSNs are one of the most viable sources for researchers to explore and analyze users' behaviors. Also, most applications that utilize user behaviors for various downstream tasks rely heavily on the observed data. For instance, consider the task of recommender systems. To ensure that the recommender system recommends relevant items to the user, the model must capture the user's behavior based on past interactions. Recently, some concentrated efforts have been dedicated to characterizing the evolution of users' activities and profiling users' behavior~\cite{antelmi2019characterizing,cheng2021causal,karami2021profiling}. However, these works overlook the concept of \textit{participation inequality} on social media. Participation inequality, also known as the 90-9-1 rule by web usability experts~\cite{nielsen}, states that various groups of people have different levels of participation. Based on this rule, the users are categorized into three types: lurkers, engagers, and contributors. The lurkers are users who only consume content without participating in content creation and comprise 90\% of OSN users. The engagers contribute sparingly to content creation and include 9\% of users. Finally, the contributors are those responsible for creating most of the content on social media and comprise 1\% of the OSN users.
As mentioned earlier, the current behavioral analysis models rely heavily on observational data. Furthermore, the majority of the observational data is created by the contributors, making these models biased toward the views of these users~\cite{gong2015characterizing}. Although effective, such models may not be generalizable for the lurkers and engagers groups. As a result, the population-level interests and sentiments are more likely to be misrepresented while utilizing these biased models. To reduce this bias, we focus on the representation of the engagers group since they have a higher observed presence when compared to the lurkers and, simultaneously, consisting of a larger population than the contributors.
 
While predicting population-level interests for groups such as the engagers, it is crucial to understand the engagers' behavior. This process can be decomposed into two steps:~(1)~estimate the user's exposure to different topics and ~(2)~interpret the reason behind their decision on engaging or not engaging with these topics or posts. In this work, we will concentrate on the first step. But, in total, understanding this complex mechanism can be highly beneficial to machine learning models. For instance, identifying the engagers' interests and disinterests can help the models recommend relevant topics. Also, in fake news classification systems, if an engager checks the veracity of a post before reacting to it, that user can be tagged as a reliable user and help improve the fake news detection~\cite{voicesilence}. When a user responds to a specific topic, it implies that the user was exposed to that topic. However, the lack of engagement would not guarantee that the user was not exposed to that post and it can occur due to various reasons. For example, (1)~a user might be a lurker and only consumes content~\cite{weeks2017online}, (2)~a user might fear the social isolation that follows by voicing their opinions on sensitive topics (the spiral of silence theory)~\cite{hampton2014social}, (3)~a user's interest may not align with the content's topic~\cite{gong2015characterizing}, and (4)~a user might doubt the veracity of the post~\cite{shu2020combating}.

Traditional behavioral analysis methods utilize an activity matrix to identify the exposure of a topic to a user~\cite{voicesilence}. These matrices ($\mathcal{A}$) are binary in nature where a value of $\mathcal{A}_{i,j}=1$ indicates that user $i$ interacted with topic $j$. However, these matrices suffer from two problems: (1)~almost 90\% of the observations in the matrices ($\mathcal{A}_{i,j}=1$) are due to the contributors' activities~\cite{antelmi2019characterizing}. With them consisting only 1\% of all the users, any analysis on these matrices would be biased towards the views of this group;
and (2)~for every unobserved user-topic interaction, the traditional systems assume that the user was never exposed to that topic, which may not be true.
\begin{figure}[ht]
  \centering
  \includegraphics[width=0.8\textwidth]{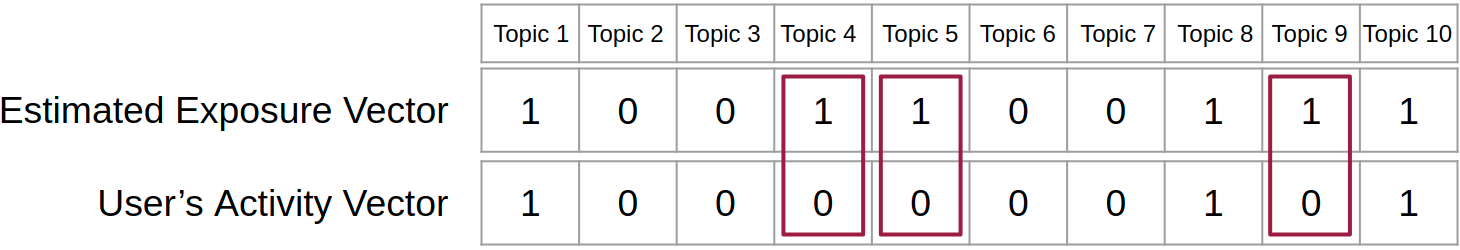}
  \caption{The exposure vector shows the topics that the user was exposed to while the activity vector denotes her interaction with a topic. Topics 4, 5, and 9 are those that the user chooses not to engage with even though she was exposed to.}
  \label{fig::exposureVSactivity} 
\end{figure} 
For instance, in Fig.\ref{fig::exposureVSactivity} we can observe the user's activity vector which indicates the topics a user interacted with, and an exposure vector that indicates the topics a user was exposed to. As seen from the example, although the user was exposed to Topics 4, 5, and 9, the user did not perform any interactions with those topics. The underlying reason behind not interacting with these topics can be any of the reasons listed above. 
To overcome this problem, in this work, we investigate how to estimate the topic exposure for behavioral analysis. We aim to model the exposure matrix for the engagers by looking into the behavior of similar users in terms of their activity and profile information.
The main contribution of this work is three-fold:
\begin{itemize}
    \item To the best of our knowledge, this work is the first attempt in estimating the topic exposure for engagers; 
    \item We conduct experiments to show the effectiveness of the proposed method;
    \item Although the focus of this paper is on the engagers, we extend a benchmark dataset to represent lurkers, engagers, and contributors which can be utilized for generalized user behavioral analysis.
\end{itemize}

\section{Problem Statement}
Let $[\mathcal{U}]_{m \times n} = \{(\mathbf{u}_1, y_1), (\mathbf{u}_2, y_2), ..., (\mathbf{u}_m, y_m)\}$ denote a set of $m$ users with $y_i$ showing whether the user is an engager ($y_i=0$) or a contributor ($y_i=1$). Lets also denote user-topic activity matrix with $[\mathcal{A}]_{m \times t} = \{\mathbf{a}_1, \mathbf{a}_2, ..., \mathbf{a}_m\}$ and user-profile matrix with $[\mathcal{P}]_{m \times (n-t)} = \{\mathbf{p}_1, \mathbf{p}_2, ..., \mathbf{p}_m\}$, where $t$ is the number of topics. Each $\mathbf{u}_i = [\mathbf{a}_i||\mathbf{p}_i]$ is a vector with size $n$ that includes user information by concatenating the user-topic activity vector ($\mathbf{a}_i$) and user-profile vector ($\mathbf{p}_i$) in which the details of these vectors is provided in the following sections. The problem is formally defined as follows:

\begin{center}
\fbox{\parbox[c]{.93\linewidth}{\textbf{Definition (Topic Exposure Matrix Estimator).} Given the $[\mathcal{U}]$ matrix comprising of the activity $[\mathcal{A}]$ and profile $[\mathcal{P}]$ information of the engagers and contributors, the goal is to estimate the exposure matrix for engagers, $[\mathcal{E}|y=0]$, which shows what topics were exposed to the engagers. }}
\end{center}

\section{Proposed Model}
\label{sec::proposed}

\begin{wraptable}{r}{7.1cm}
  \centering
  \includegraphics[width=0.58\textwidth]{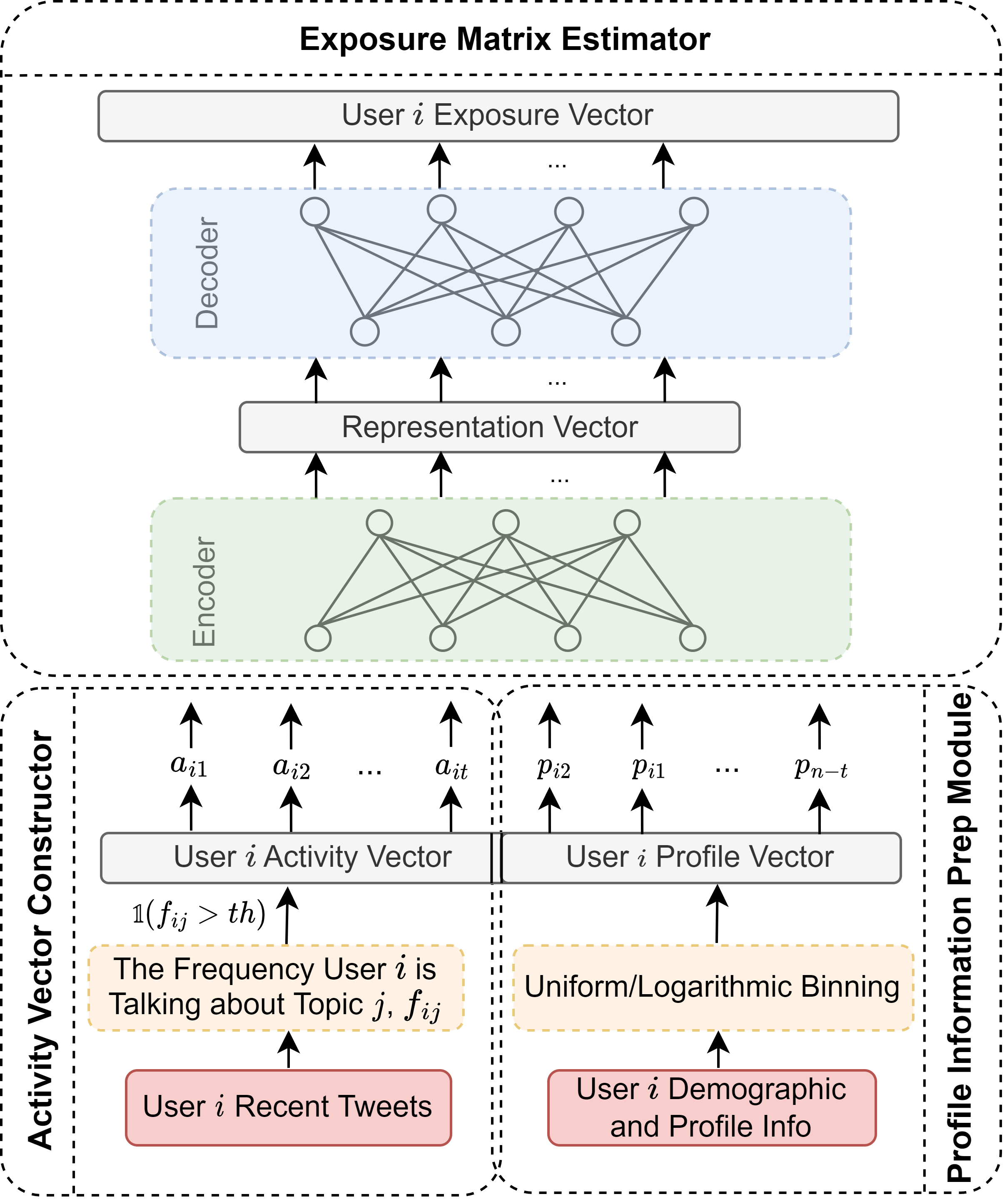}
  \caption{The three components of the proposed model: (1)~Activity vector constructor, (2)~Profile information prep module, and (3)~Exposure matrix estimator.}
  \label{fig::proposed} 
\end{wraptable}
We divide the users into three different categories: Lurkers, Engagers, and Contributors. For our analysis, we eliminate lurkers from the data due to the following reasons: (1)~They generate no network value on social media, (2)~We do not have access to the users' login information which is the time they enter the platform until they leave the application, and (3)~The time spent by the user on each content posted on their timeline is unknown such as the duration of the pause before scrolling from the displayed post. We concentrate on engagers to estimate their topic exposure since they are an under-represented group but still leave their footprint on social media. The general framework of the proposed model is shown in Fig.\ref{fig::proposed}. The proposed model
has three major components: (1)~Activity vector constructor, (2)~Profile information prep module, and (3)~Exposure matrix estimator. The model is flexible in the algorithm used to extract the topics.
Based on the application and whether it is important to know the explicit topics or latent topics, we can use methods as simple as a bag of words or TF-IDF or utilize topic modeling approaches such as Latent Dirichlet Allocation (LDA)~\cite{blei2003latent}, or apply recent methods like BERTopic~\cite{grootendorst2020bertopic}.
We describe next in detail each of the model's components and discuss how they are integrated into a framework that predicts the exposure matrix for the engagers.

\subsection{Activity Vector Constructor}
In creating the user-topic activity matrix, we look into the tweets of the users which explicitly reflects their opinion on the topics. To this end, let us introduce a set of binary variables $a_{ij}$, which indicates the engagement of user $i$ on topic $j$. The $a_{ij}$ values are derived by binarizing the $f_{ij}$ values:
\begin{equation}
\begin{aligned}
    a_{ij}= 
\begin{cases}
    1,& \text{if } f_{ij}>th\\
    0,              & \text{if } f_{ij} = 0\
\end{cases},
\label{eq:utamat}
\end{aligned}
\end{equation}
\noindent
where $f_{ij}$ shows the frequency a user $i$ talked about topic $j$ out of the recent tweets of that user. For example, if 15\% of the $i$'s recent tweets is about topic $j$ then $f_{ij} = 0.15$. Moreover, the threshold, $0 \leq th \leq 1$, is a predefined value showing the user's engagement intensity.
In other words, if user $i$ reacts to topic $j$ more than a specified threshold, $f_{ij}>th$, then we have an indication that $i$ is engaged with $j$ or $a_{ij} = 1$. On the other hand, if $i$ has never tweeted about $j$ (i.e.,~$a_{ij} = 0$), then it is uncertain whether the observed 0 is due to the lack of exposure to that topic or whether the user was exposed but chose not to speak about that topic. We denote the user-topic activity matrix, $[\mathcal{A}]_{m \times t}=[a_{ij}]$, as a binary matrix of $m$ users and $t$ topics.

\subsection{Profile Information Prep Module}
To create the binary matrix of the user profile, we extracted the explicit and demographic information of the users such as age, gender, whether they are verified users, whether they are an organization, the number of days the user has been registered to the platform, and their followers and friends count. Categorical features were transformed into binary based on the number of available categories. We used uniform binning to discretize numerical values with normal distribution such as the registered time and logarithmic binning for values with a power-law distribution like friends and followers count. This would also help in handling outliers as well as minimizing the effect of indifferent gaps in the value spread. For example, a user with 100 followers is not that different from a user with 120 followers and a user with millions of followers is rare (i.e., an outlier).

\subsection{Exposure Matrix Estimator}
To estimate the exposure matrix, we utilize an encoder-decoder with two hidden layers. The networks are trained using Binary Cross Entropy loss function:
\begin{equation}
    \mathcal{L}_{BCE} = -\frac{1}{M} \sum_{i=1}^{M} (\mathbf{e_i} \log (\mathbf{\hat{e}}_i) + (1-\mathbf{e_i}) \log (1-\mathbf{\hat{e}}_i)),
    \label{eq:ce_loss}
\end{equation}
where $M$ is the output size (number of topics), $\mathbf{e_i}$ and $\mathbf{\hat{e_i}}$ are the ground truth and the prediction of the exposure matrix in that batch, respectively.



\section{Experiments}
We conduct experiments to evaluate the effectiveness of our method in estimating the topic exposure matrix. We propose two research questions:

\begin{enumerate}[leftmargin=.5in, label={(Q\arabic*)}]
        \item How well do the proposed method and other similar methods estimate the user-topic exposure matrix?
        \item Will adding users' demographic and profile information help in improving the estimation?
\end{enumerate}
We will start by introducing the datasets and the baseline methods used for answering these questions and then discuss the experimental results.
\subsection{Datasets}
We used two datasets from the FakeNewsNet repository as the seed dataset to conduct our experiments: PolitiFact and GossipCop~\cite{shu2020fakenewsnet}. The reason we choose these datasets is that the datasets also include the user's demographic features such as age and gender as well as the user's explicit information such as the number of followers.
As for the users' tweets, we collected 200 recent posts of each user identified in the two datasets. To identify three groups of lurkers, engagers, and contributors, we calculate the average number of activities per day. 
If a user does not engage in any activity or near to zero engagement, that user is classified as a lurker and is eliminated from our experiments.

In this work, two assumptions have been considered: (1)~we assume that similar users would have similar behavior. This assumption has been well justified in many classical methods in machine learning such as nearest neighbor classification; (2)~since the contributors would often engage with the topics that they are exposed to due to different reasons including individual's personality and the communities' pro-sharing norms~\cite{sun2014understanding}, their activity vector can serve as an approximated proxy of their exposure. The goal is to propose a model that approximates the exposure matrix for the engagers, such that we would be able to identify the topics that these users choose not to engage with (Fig.\ref{fig::exposureVSactivity}).

\begin{wraptable}{r}{5.6cm}
\centering
\footnotesize
\caption{Statistics of the datasets.}
\label{tab:stat}
\begin{tabular}{lcc}
\hline
            \textbf{User Type}       & \textbf{Gossipcop} & \textbf{Politifact} \\
                   \hline
Lurkers & 436 & 1,099\\
Engagers &    2,575    & 6,138       \\
Mock Engagers & 4,413 & 15,699 \\
Contributors   &   15,120    & 62,162      \\
\hline
Total              &   22,544    & 85,098     \\
\hline
\end{tabular}
\end{wraptable}

Since we do not have access to the actual exposure matrix for the engagers (i.e. no ground truth), we create a set of engages, named mock engagers, out of the contributors' data. In this case, based on our second assumption, the actual activity vector of the contributor can act as the ground truth for the mock engager. To this end, we eliminate tweets randomly per day from the contributors' recent tweets such that the average number of tweets per day becomes equal to or less than the threshold set for the engagers. We set the thresholds for the average number of activities per day in creating the lurkers and engagers to 0.005 and 0.1, respectively, such that it approximately follows the 90-9-1 rule~\cite{nielsen}. Statistics of the final created datasets are summarized in Table~\ref{tab:stat}. 

\subsection{Experimental Design}
\begin{figure}[ht]
  \centering
  \includegraphics[width=0.8\textwidth]{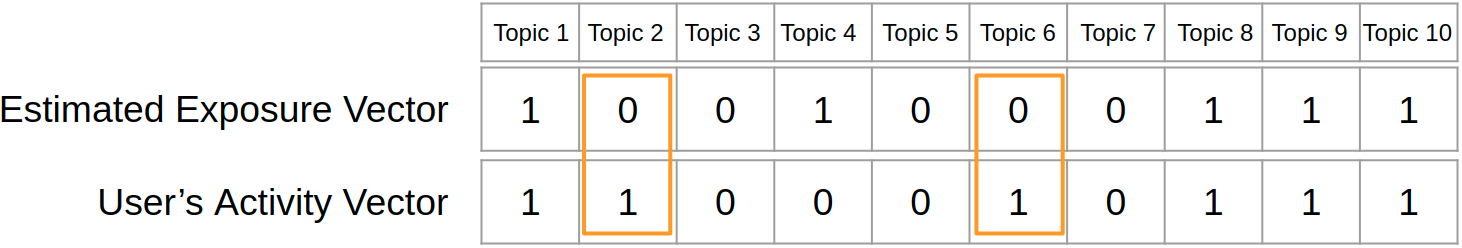}
  \caption{We predicted that the user was not exposed to topics 2 and 6, while in reality she already talked about them, which shows a contradiction. In estimating the exposure vector, we want to avoid cases where the value in the estimated exposure vector is 0 while its corresponding value in the activity vector is 1.}
  \label{fig::exposureVSactivity2} 
\end{figure}

Several methods were used to estimate the exposure matrix and evaluate the effectiveness of the proposed method. We will introduce these methods below:
\begin{itemize}
    \item \textbf{Nearest Neighbor Matching:}~We start with the simplest method where for estimating the exposure vector of each [mock] engager we choose their nearest neighbor match from the contributors' group. However, this might not be a good estimate as this might pick up the nearest contributor in which the user would not talk about the same topics that the engager does~(Fig.\ref{fig::exposureVSactivity2}). We are mostly looking for cases in which the engager and the contributor both talked about the same topics, while the contributor might talk about other topics as well. In other words, we want to minimize the cases when we have zero value for a topic in the estimated exposure vector while its corresponding activity value is one~(Conditional Nearest Neighbor Matching).
    \item \textbf{Hierarchical Poisson Factorization:} Originally proposed by Gopalan et al.~\cite{gopalan2015scalable}, Wang et al.~\cite{wang2020causal} utilized this method for movie recommender systems. The method first uses Poisson Factorization~(PF) to fit a model to the observed user-movie ratings (i.e. the user-topic activity vector, in our case). Next, it reconstructs the exposure matrix from the PF fit which accounts for unobserved confounders. More formally, lets assume that $\pi_u \stackrel{\scalebox{.5}{$
   i.i.d$}}{\sim} \text{Gamma}(c_1,c_2)$ captures user preferences and $\lambda_t \stackrel{\scalebox{.5}{$
   i.i.d$}}{\sim} \text{Gamma}(c_3,c_4)$ captures topic attributes. Then PF assumes the data come from the following process:
    \begin{equation}
        a_{ut}|\pi_u, \lambda_t \;\; \sim \;\; \text{Poisson}(\pi_u^T\lambda_t) \quad \forall u,t,
    \label{eq:pf1}
    \end{equation}
the exposure matrix is reconstructed from the PF fit by:
    \begin{equation}
        \hat{e}_{ut}=\mathbb{E}_{\text{PF}}[\pi_u^T\lambda_t|\mathbf{a}].
    \label{eq:pf2}
    \end{equation}
    
\end{itemize}

For the experiments, we used TF-IDF to assign weights to the topics and extracted the top 100 ones discussed by the users in each dataset. As for the user-topic activity matrix we set the threshold, $th$, defined in equation~\ref{eq:utamat} to zero. So, if user $i$ mentions topic $j$ in one or more posts, ($f_{ij}>0$), $a_{ij}$ is set to 1, otherwise it is 0. We used two hidden layers of 64 and 32 neurons for the encoder-decoder and trained it using an Adam optimizer with a learning rate of 1e-4.

\subsection{Discussion and Experimental Result}

To evaluate the estimated exposure matrix from the ground truth, we utilize the normalized $L_1\text{-norm}$ [dis]similarity measure:
\begin{equation}
    L_1\text{-norm} = \frac{1}{m\cdot t}\sum_{i=1}^m \sum_{j=1}^t |e_{ij} -  \hat{e}_{ij}|,
    \label{eq:l1_norm}
\end{equation}
\noindent
where $m$ is the number of users, $t$ is the number of topics, $e_{ij}$ shows the entries in the exposed matrix, and $\hat{e}_{ij}$ is the corresponding estimated value. The lower the value of this measure, the more similar the estimated vector is to the actual one. Table~\ref{tab:result} shows the error rate for the baselines and the proposed method.

\begin{table}[h]
\centering
\small
\caption{$L_1$-norm distance between the actual and the estimated exposure matrix with and without considering the profile information (lower values are better). Hierarchical Poisson Factorization does not work with profile information.}
\label{tab:result}
\begin{tabular}{lcc}
\toprule
\multicolumn{3}{c}{\textbf{Without Profile Information}} \\
\hline
Model & Gossipcop & Politifact \\
\hline
Nearest Neighbor Matching & 0.33 $\pm$ 0.11 & 0.33 $\pm$ 0.10\\
Conditional Nearest Neighbor  &   0.31 $\pm$ 0.10     &   0.31 $\pm$ 0.90     \\
Hierarchical Poisson Factorization & 0.40 $\pm$ 0.13 & 0.43 $\pm$ 0.14\\
\hline
Proposed Model  & \textbf{0.26 $\pm$ 0.09} & \textbf{0.25 $\pm$ 0.08} \\
\hline
\midrule
\multicolumn{3}{c}{\textbf{With Profile Information}} \\
\hline
Model & Gossipcop & Politifact \\
\hline
Nearest Neighbor Matching &  0.18 $\pm$ 0.09 & 0.18 $\pm$ 0.10 \\
Conditional Nearest Neighbor  & 0.16 $\pm$ 0.08 & 
0.17 $\pm$ 0.09
\\
\hline
Proposed Model  & \textbf{0.07 $\pm$ 0.05}  &  \textbf{0.06 $\pm$ 0.05}\\

\bottomrule
\end{tabular}
\end{table}

\textbf{Q1.} As expected, the conditional nearest neighbor works better than the pure nearest neighbor matching as it minimizes the situation in which the estimated exposed value for a topic is zero while the corresponding value in the activity vector is one. Additionally, the proposed method has shown superior results over other methods providing evidence that our model can best estimate which topics are exposed to users. 

\textbf{Q2.} On the other hand, adding profile information in reconstructing the estimated exposed matrix, would improve the models' performance significantly. From these results, it is evident that social media platforms such as Twitter would display topics to users not only based on the previous activities of the users but also with respect to the users' profile information. In this case, the proposed method was able to outperform the other models as well. The Hierarchical Poisson Factorization cannot be used with profile information as it fits the model by only considering the user-topic matrix.


\section{Related Work}
The proposed method spans the subject domains of user behavioral analysis and online participation. Users on social media are able to engage in various online social activities such as posting, browsing, and social networking. Understanding and analyzing their behavior is important, particularly for service providers and researchers. 
These available data on social media can be divided into four categories: (i)~user's profile information, (ii)~user's activity, (iii)~user's network connectivity, and (iv)~user's generated content. All these categories are important in analyzing the user's behavior and characteristics. In the field of social media behavioral analysis, researchers have utilized different conjunctions of these categories of data~\cite{antelmi2019characterizing,cheng2021causal,tahora2019bot,karami2021profiling}. However, little attention has been given to the concept of participation inequality. More precisely, these methods do not distinguish between different types of users in terms of the amount of their created content and they commonly interpret the users' silence as a lack of engagement~\cite{voicesilence}. 
In the field of psychology, behavioral, and social science, there is a wide range of studies dedicated to extracting factors that drive user participation as well as lurking behavior in online social communities~\cite{sun2014understanding}. These behavioral factors can be classified into three major categories: (1)~individual-level which takes into account the intrinsic and personal characteristics of the users such as their personality traits, (2)~community-level which is related to the social aspect and the nature of the online communities, and (3)~environmental-level which looks into the extrinsic environmental influence such as the platform's functionalities, characteristics, and development process. In other words, with suitable and safe infrastructure, due to the users' personal characteristics as well as the reciprocity factor in OSNs, the users have the confidence to contribute to the online content as well as perceiving that their posts are useful for the community and receive by a lot of people.   Moreover, these studies discuss that lurking behavior is actually normal, positive, active, and a valuable form of online behavior~\cite{chen2019seeking,edelmann2013reviewing}. By providing an explanation of the users' silence we can design strategies to encourage posting~\cite{sun2014understanding} or devise unbiased social methods~\cite{liu2016characterizing} such as more accurate election predictors~\cite{gayo2012wanted} and fact-finding algorithms~\cite{voicesilence}.

\section{Conclusion and Future Work}
To the best of our knowledge, this is the first work that attempts to estimate the topic exposure matrix for engagers which are often misrepresented in behavioral analysis research. Their voices are ignored mostly due to their little contribution to the online content, even though they are large in number. This paper would help in identifying the topics these users choose not to engage with. Later, providing explanations of their silence and interpreting their behavior can help the researchers in designing unbiased social methods including social sensing.
%
%
%
\bibliographystyle{splncs04}
\bibliography{mybibliography}
%




\end{document}